\documentclass[conference,twosided,10pt]{IEEEtran}
\usepackage[utf8]{inputenc}

\usepackage{hyperref}
\title{Where are the Teeth of the Fourth Power?}

\author{
  \IEEEauthorblockN{Lutz Stra$\ss$burger}
  \IEEEauthorblockA{Inria Saclay ~\&~ École Polytechnique\\
  }
}

\begin{document}

\maketitle

Our modern democracies are based on the separation of powers into legislative, executive, and judicial branches. In this setting, the mass media (such as the press, television, and radio) represents a fourth, virtual pillar, which, without legally anchored power, is intended to exercise a control function over the three state powers by means of truthful reporting and mediation of public opinion in order to prevent abuse of power.\footnote{In English, the term ``Fourth Power'' is less common than in other European languages, e.g., German (Vierte Gewalt), Italian (quarto potere), Spanish (cuarto poder),
French (quatrième pouvoir), or Polish (Czwarta Władza).}
\looseness=-1

Indeed, in 1891, Oscar Wilde complained \emph{``In old days men had the rack. Now they have the Press. [...] We are dominated by Journalism.''}~\cite{wilde:journalism}
And in the 20th century, one of the most well-known successes of the \emph{Fourth Power} is the coverage of the Watergate scandal in the 1970s, that eventually led to the resignation of US-President Richard Nixon.

Fast-forward to our century. In 2010, WikiLeaks and its media partners published the \emph{Iraq War Logs} documenting war crimes of the US military;\footnote{\url{https://wikileaks.org/irq/}} in 2013 several newspapers worldwide reported on the largest surveillance scandal, uncovered by wistleblower \emph{Edward Snowden}; and in 2018, European newspapers in several countries published the \emph{CumEx-Files}, uncovering a tax fraud via a dividend arbitrage mechanism between banks and financial investors, where 150 billion euros have been stolen from the national tax authorities of several European states.\footnote{\url{https://correctiv.org/en/latest-stories/2021/10/21/cumex-files-2/}}
However, in none of these cases have responsible politicians been held accountable. In fact, in the case of the \emph{Iraq War Logs}, the people who uncovered the war crimes went to jail, but not the people who committed the crimes. In the case of the illegal surveillance made public by Edward Snowden, lawmakers in various countries changed the laws so that the former illegal surveillance is now legal. And in the case of CumEx, it seems that some of the involved politicians are now in more powerful positions. 
What did change? Why did the press lose its bite?

An easy answer could be the argument that our demands and expectations are simply too high and that it was never a task of the media to protect the democracy, as the media have their own economical interests, as in our times, ``generating clicks'' is more important than ``truthful reporting''. This problem has already been observed in the 1980s~\cite{herman:chomsky:consent}. But even though this is still an underinvestigated topic, it is not what I want to discuss here.\looseness=-1

After all, observe that the examples I mentioned above have all been properly reported by the media. The publications did just not have the consequences, that a naive law abiding citizen might have expected.

Why?

The problem that I want to address here is not one of political science or social science, but one of computer science.\looseness=-1


In previous centuries, people were usually reading one or maybe two daily newspapers, and these newspapers were providing an all-encompassing news coverage. That means that a person reading the whole paper could assume to more or less know what was going on in the world.
Today, the situation is different. Via the internet, we have access to all the articles in all the newspapers, and it is impossible to read all of them.

How to choose which ones to read?

This choice is made with the help of algorithms. Either we use them to search for news, or we even allow them to suggest a list of news to read. This has certain consequences,  and three of them I would like to discuss here.

\medskip

The first is the creation of \emph{Filter Bubbles}~\cite{pariser:filter}: As the algorithms select what we like to see, and as we tend to befriend people on social media who share our opinions, we eventually only see news that support our own viewpoints.

Coming back to the comparison with newspapers in the 20th century: if two people were reading two different newspapers, they probably had still access to the same news. But today, it is very unlikely that you and your neighbor are getting the same news. In fact, even if you are looking for the same thing with the same search engine, you might get very different results~\cite{pariser:filter}.
This means, that, \emph{a priori}, we cannot know if we are reading the same news as everyone else.

This phenomenon comes now slowly into the public awareness and computer scientists, including logicians begin to investigate it~\cite{venema-los:MajorityIllusions,BCRS:diffusion}, but the potential impact of filter bubbles to the functioning of our society is still not properly understood.

\medskip

The second consequence poses a danger that is much more severe than filter bubbles and that the public seems totally unaware of: whoever controls the algorithms can control our political opinions and viewpoints. The phenomenon of filter bubbles occurs under the assumption that the algorithms are neutral and simply give us what we like. But what guarantees that the selection algorithms are neutral and do not slowly guide us towards other opinions? Is there any way to prevent this from happening, or at least observe it when it is happening?\looseness=-1

In 2018 the \emph{Cambridge Analytica scandal}~\cite{wylie:mindf*ck} was in the news, as the company had collected private data from 87 million Facebook users without their consent, in order to target them with tailored adds for various political campaigns, including Brexit and the 2016 presidential election in the US. 

This led to a public discussion of various privacy issues. And even though it is certainly a positive effect when the need for privacy and data protection is in the public awareness, it is important to observe that the lack of privacy was never the problem in the Cambridge Analytica scandal. All the private information was given voluntarily to Facebook by its users, and the scandal was focusing on the way how Cambridge Analytica got the data from Facebook. Now, everybody was talking about how Cambridge Analytica was analyzing the data of a fraction of Facebook users to target them with personalized messages. But why is nobody talking about the fact that Facebook has all the data of all Facebook users and could potentially target all of them with personalized messages?\looseness=-1

In fact, when it comes to targeted advertisement, people are aware of the problems; and the dangers and mechanisms are beginning to be investigated by computer scientists (e.g.~\cite{goga:children}), and this also includes political ads (e.g.~\cite{goga:policy}).

But what about targeted news?

As a matter of fact, Facebook is already targeting each Facebook user with personalized news, as this is exactly what the selection algorithms that I mentioned above are doing. 

For a long time now, psychologists and behavioral scientists have a very good understanding of how easy it is to influence and manipulate people~(e.g.~\cite{cialdini:influence,kahneman:fastslow}
); and recent research also suggests that the order in which we receive information influences our behaviors and opinions~\cite{cialdini:presuation}.\footnote{The basic principles were already known in the 1980s, as this short clip from the BBC Comedy Show \emph{Yes, Prime Minister} shows: \url{https://www.youtube.com/watch?v=ahgjEjJkZks}}

And who determines the order in which we read our news?

It should be clear that we cannot simply trust large companies like Facebook or Google or Apple that they do not abuse their power.\footnote{If we learned something from human history, then that
whenever something can be done, then it will be done.}
Neither can we trust the people that they would notice manipulations and complain. In fact successful manipulation is undetectable by the person being manipulated.

Furthermore, for swaying an election only a small fraction of voters has to be influenced, and companies like Facebook have the information on which fraction that is.

\medskip

The third consequence that I would like to discuss here is a bit more subtle. Let us return to the starting point of this essay: Could it be that in the case of the Watergate scandal, the press succeeded in keeping the attention of the people long enough on the scandal so that eventually the public created enough pressure to cause the resignation of Nixon? And could it be that in the cases of Edward Snowden, WikiLeaks, and CumEx, the attention of the public has been drawn away too quickly so that not enough public pressure was created? 
Of course, it could all be a cognitive illusion and there are completely different explanations. However, my point is that even if the selection algorithms are neutral, they might simply draw us away from news that give us a bad feeling towards news that make us feel good, simply because that is how the human brain is directing our behavior. This effect is increased through the creation of so called \emph{fake news} by dubious interest groups. 

Since the beginning of the media, governments have understood their power. Already Nazi-Germany used very successfully radio (and later also TV) for their propaganda. Dictatorships always first control the media, in particular, what information they provide, and more importantly, what information they hide. That is why in our democracies we have laws ensuring the \emph{freedom of the press}. However, if we let algorithms decide what information we consume, the government does not need to restrict our information access. We do it ourselves.

\medskip

Let me summarize the effects of the use of selection algorithms for our news consumption:
\begin{enumerate}
\item creation of filter bubbles,
\item potential manipulation of our opinions and behaviors by whoever controls these algorithms, and
\item quick distraction from uncomfortable news.
\end{enumerate}
Unfortunately, these selection (and search) algorithms are trade secrets of large companies like Facebook, Google, or Apple. This is not going to change anytime soon, and the situation is only getting worse when some form of AI is used for that. However, for guarantying the proper functioning of our society, we need to understand how these algorithm operate. This means we have  to treat them as \emph{black boxes}, and the first and foremost task is to find a way to measure the three effects mentioned above. 

This is quite a challenge, as none of these effects is detectable by the individual user who is affected by them. For this reason I think that a proper quantitative analysis can only be achieved by a combined effort of computer scientists and linguists and social scientists.



\bibliographystyle{plain}
\bibliography{lutzrefs}
\end{document}